\documentclass[conference]{IEEEtran}
\usepackage{graphicx} 
\usepackage{xcolor}
\usepackage{array}
\usepackage{float}
\usepackage{setspace} 
\usepackage{textcomp}
\usepackage{xcolor}
\usepackage{url}
\usepackage{threeparttable}
\usepackage{ragged2e}
\usepackage{multirow}
\usepackage{diagbox}
\usepackage[backend=biber, bibstyle=ieee, citestyle=numeric, date=year, sorting=nyt, doi=true]{biblatex} 
\usepackage{amsmath,amssymb,amsfonts}
\usepackage{fancyhdr}
\def\BibTeX{{\rm B\kern-.05em{\sc i\kern-.025em b}\kern-.08em
    T\kern-.1667em\lower.7ex\hbox{E}\kern-.125emX}}
\title{Bitcoin Price Prediction using Machine Learning and Combinatorial Fusion Analysis}
\author{\IEEEauthorblockN{Yuanhong Wu\IEEEauthorrefmark{1},
Wei Ye\IEEEauthorrefmark{2},
Jingyan Xu\IEEEauthorrefmark{1},~\IEEEmembership{Student~Member,~IEEE} and
D. Frank Hsu\IEEEauthorrefmark{1},~\IEEEmembership{Senior~Life~Member,~IEEE}}
\IEEEauthorblockA{\IEEEauthorrefmark{1} Laboratory of Informatics and Data Mining, \\
Department of Computer and Information Science, Fordham University, New York, NY 10023, USA.}
\IEEEauthorblockA{\IEEEauthorrefmark{2} Department of Economics, Fordham University, New York, NY 10023, USA.}
\thanks{
Corresponding authors: E-mail: \{ywu463; hsu\}@fordham.edu}}

\IEEEoverridecommandlockouts
\begin{document}

\maketitle
\pagestyle{fancy}
\fancyhf{}
\fancyhead[L]{Published as a conference paper at 2025 IEEE Conference on Artificial Intelligence (IEEE CAI)}
\renewcommand{\headrulewidth}{0.4pt}

\begin{abstract}
In this work, we propose to apply a new model fusion and learning paradigm, known as Combinatorial Fusion Analysis (CFA), to the field of Bitcoin price prediction. Price prediction of financial product has always been a big topic in finance, as the successful prediction of the price can yield significant profit. Every machine learning model has its own strength and weakness, which hinders progress toward robustness. CFA has been used to enhance models by leveraging rank-score characteristic (RSC) function and cognitive diversity in the combination of a moderate set of diverse and relatively well-performed models. Our method utilizes both score and rank combinations as well as other weighted combination techniques. Key metrics such as RMSE and MAPE are used to evaluate our methodology performance. Our proposal presents a notable MAPE performance of 0.19\%. The proposed method greatly improves upon individual model performance, as well as outperforms other Bitcoin price prediction models. 
\end{abstract}


\section{Introduction}
Bitcoin, introduced by Satoshi Nakamoto, is a peer-to-peer cash payment system designed to eliminate centralization and serve as a next-generation currency \cite{nakamoto2008bitcoin}. It leverages blockchain technology, the proof-of-work mechanism, and digital signatures to address challenges like double spending and the Byzantine Generals Problem. However, due to the volatility of Bitcoin's price, it is not well-suited as a currency, particularly as a medium of exchange. Its volatility may occur during the path towards equilibrium point. However, as the future is never certain, the market may not price bitcoin correctly \cite{catalini2022some}.


With the approval of spot Bitcoin exchange-traded products by the SEC on January 14, 2024 and the 2024 U.S. presidential election race projection, Bitcoin prices reached a historical high of over \$94,000. According to CoinMarketCap, Bitcoin's market capitalization stands at \$1.82 trillion, with a 24-hour trading volume exceeding \$66.43 billion. Bitcoin is now accessible not only to large investment firms such as Jump Trading and BlackRock but also to the general public, as the public can purchase Bitcoin through centralized exchanges (CEXs) like Coinbase and Binance or being swapped via decentralized exchanges (DEXs) like Uniswap. Once invested in Bitcoin, an inevitable question arises: what will the price of Bitcoin be the next day? Should one buy more, hold their current position, or sell the Bitcoin in their portfolio? To address this, a robust algorithm capable of predicting Bitcoin's price is essential. In the regard, we make use of Combinatorial Fusion Analysis, a recently developed computational learning and modeling paradigm,  to predict bitcoin price. 

\subsection{Machine Learning Models}
There have been extensive works on Bitcoin price prediction. 
One category focuses on predicting the direction of Bitcoin price movements (e.g., up or down), which frames the problem as a classification task. Another one aims to forecast Bitcoin prices directly on a testing dataset, which treats it as a regression problem.

Previous studies have explored Bitcoin price prediction, with some of them focusing on forecasting the next day's Bitcoin price. For instance, the performance of time series models (ARIMA) and neural networks (Neural Network Autoregression) was compared, and it demonstrated that time series models are more adept at capturing Bitcoin's volatile price changes compared to neural networks \cite{munim2019next}. Other studies have investigated traditional machine learning algorithms, such as Random Forest (RF), Support Vector Machine (SVM), Logistic Regression, and XGBoost, as well as other deep learning methods.

Random Forest has also been shown to perform well when Bitcoin prices are below \$60,000, but struggle when prices exceed this threshold \cite{chen2023analysis}.  In addition, statistical methods sometimes outperform more complex machine learning models for daily Bitcoin price prediction. For instance, a statistical method has achieved an accuracy rate of 66\% for this task \cite{chen2020bitcoin}. In another work, researchers finds that LSTM outperforms CNN for Bitcoin price prediction, with a MAPE of 0.196\%, which highlights LSTM's ability to capture temporal dependencies in time-series data \cite{singh2024analyzing}.  Different deep learning models have unique strengths for different tasks: LSTM excels in regression problems, while Deep Neural Networks (DNNs) are better suited for classification problems \cite{ji2019comparative}.

At the high-frequency data level, such as 5-minute time intervals, GRU has shown superior predictive power. For instance, a previous work achieves a Mean Squared Error (MSE) of 0.00002 using GRU, which illustrates its effectiveness in handling short-term, high-frequency Bitcoin price fluctuations \cite{phaladisailoed2018machine}.

\subsection{Combinatorial Fusion Analysis}
Combinatorial Fusion Analysis (CFA), introduced by Hsu, Chung, and Kristal \cite{hsu2006combinatorial}, utilizes the rank-score characteristic (RSC) function and models' diversity strengths as weights to perform score and rank combinations \cite{hsu_kristal_schweikert_2024}. 

The basic idea of CFA is that as all algorithms and models have their each respective strengths and weaknesses, superior performance can be achieved through model fusion. CFA has been successfully applied in a variety of domain applications, including Denial of Service (DOS) attack detection \cite{owusu2023enhancing}, 
target tracking \cite{lyons}, and drug discovery \cite{jiang}.

Several applications of CFA have also been explored in finance. One prominent example is its use in portfolio management, where CFA has been shown to enhance portfolio management strategies in the U.S. stock market \cite{irukulapati2018long}. Similarly, there has also been work to apply CFA techniques, including rank combination, score combination, and rank-score combination, to rank U.S. equities \cite{jiang2023equity}. Those results demonstrate that CFA outperforms the iShares S\&P 500 over the same period in terms of Sharpe Ratio and cumulative return.

Current study is the first to apply the CFA method to facilitate the prediction of Bitcoin's next-day price. In this approach, we obtain a scoring system for the next day's price from each of the five algorithms employed. The CFA method is then utilized to combine these five scoring systems, resulting in a single combined prediction for Bitcoin's next-day price.

Our results demonstrate that the CFA method performs effectively on the testing dataset, as evidenced by a lower MAPE than other models. Furthermore, our combined approach outperforms the results reported in previous studies, highlighting the efficacy of CFA in price prediction.

Section II provides a comprehensive overview of the methodology which comprises four subsections: subsection A details the dataset processing steps; subsection B discusses the five base models employed in this study; subsection C introduces the CFA technique; and subsection D outlines the overall workflow of the methodology. Section III presents the results, comparing the performance of the proposed model with those of the base models and other existing research studies. Finally, Section IV concludes the paper, summarizes the key findings, and discusses potential directions for future research.

\section{Methodology}
\subsection{Data Preparation}
\begin{figure}[H]
    \centering
    \includegraphics[width=0.95\linewidth]{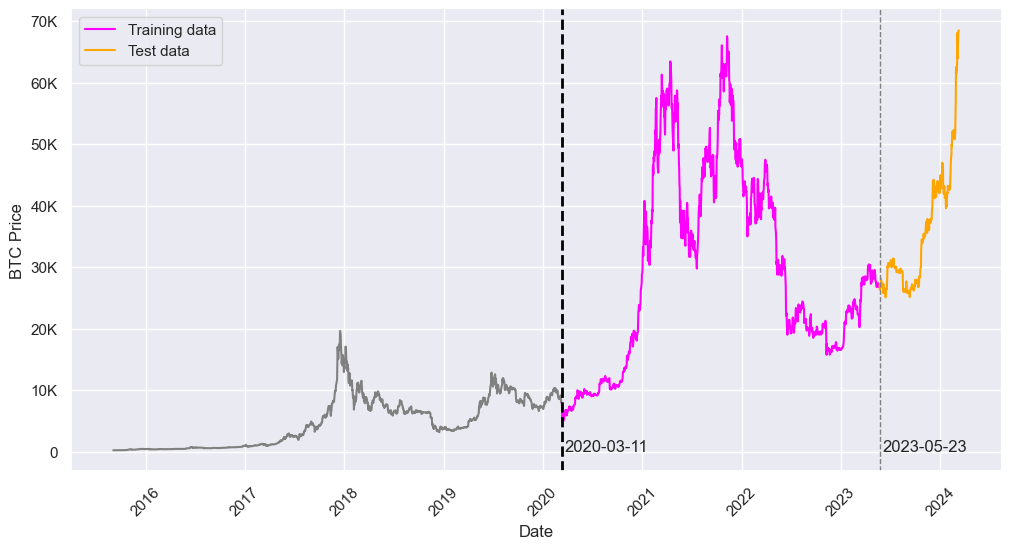}
    \caption{Historical Bitcoin price and data partitioning}
    \label{price}
\end{figure}

The dataset includes daily-level data from sources shown in Table \ref{tab:datasource}, covering the period from August 10, 2015, to March 9, 2024. Our primary goal is to predict the next day's Bitcoin price. However, the first few years' Bitcoin price series provides limited informational gain about subsequent prices because the most significant Bitcoin price gains occurred during the COVID-19 period. Therefore, we narrow our prediction task's data range to start from March 11, 2020, to better capture relevant trends. As Figure \ref{price} shows,  the dataset spans from March 11, 2020, to March 9, 2024. We partitioned the data into a training set (magenta) and a test set (orange) with an 80:20 ratio.

\begin{table}[H]
    \centering
    \small
    \caption{Data Source}
    \begin{tabular}{c|c}
    \hline
        Data &  Source\\
        \hline
        Bitcoin & Coinmetrics\\
        ETH & Coinmetrics\\
        Gold Price & Yahoo Finance\\
        Hashrate & Blockchain.com\\
        S\&P500 & Yahoo Finance \\
        Vix Index & Yahoo Finance\\
        US bond yield & Yahoo Finance\\
        Dollar Strength & Yahoo Finance\\
        Nvidia Price & Yahoo Finance\\
        Tesla Price & Yahoo Finance\\
        \hline
    \end{tabular}
    \label{tab:datasource}
\end{table}

The features selected for prediction include ETH price and gold price, along with the hash rate, which measures miners' computational power. These features have been proved effective for the task of bitcoin predictions \cite{chen2023analysis}. A higher hash rate indicates a more stable Bitcoin network. Other features include the S\&P 500 index and the VIX index, which measures market uncertainty. The US bond yield, considered a proxy for the risk-free rate, and dollar strength, which compares the US dollar to a basket of other currencies, are also included. Nvidia’s price is considered due to its role as a major chip provider, which correlates with miners' demand for ASIC hardware. Finally, Tesla's price is included as an indirect measure of Elon Musk's influence on cryptocurrencies.

Inspired by equity analyses, we also incorporate technical indicators such as the Exponential Moving Average (EMA) and the Moving Average Convergence Divergence (MACD) line.

Bitcoin is traded 24/7, whereas most of the selected features are only available on weekdays. To address this discrepancy, missing values in these features are forward-filled using the last available non-missing values. To ensure consistency, we only use daily Bitcoin price data to align with the daily frequency of the selected features. Since hash rate is updated every three days, missing values are similarly forward-filled, as hash rate changes are minimal between adjacent days.

To prepare the data for the models, we normalize each of the features to the range [0, 1]. 

\subsection{Base Models}

Five base models are used in our study. They are: SVM, Random Forest, XGBoost, CNN, and LSTM.
\subsubsection{SVM}

Support Vector Machines (SVMs) are a class of supervised learning models widely used for classification and regression tasks \cite{SVM}. The central idea of SVM is to identify a hyperplane in a high-dimensional space that best separates the data into distinct classes. This hyperplane is chosen to maximize the margin, defined as the distance between the closest data points (support vectors) of each class and the hyperplane itself. 


\subsubsection{Random Forest}

Random Forest is an ensemble learning method that builds a multitude of decision trees during training and aggregates their outputs to make robust predictions \cite{RF}. It operates by constructing each tree on a randomly selected subset of the data, both in terms of features and training samples. 


\subsubsection{XGBoost}
eXtreme Gradient Boosting (XGBoost), is a scalable and fast implementation of the gradient boosting framework widely used in supervised learning tasks \cite{Chen_XGBoost}. It builds an ensemble of decision trees sequentially, where each tree corrects the errors of the previous ones, optimizing with second-order gradients for improved performance. 


\subsubsection{CNN}

Convolutional Neural Networks (CNNs) are neural networks designed for grid-like data, such as images, mimicking how neurons in the brain respond to visual stimuli \cite{LeCun_CNN}. They consist of layers like convolutional layers, which extract features using sliding kernels, pooling layers that reduce spatial dimensions, and fully connected layers. 


\subsubsection{LSTM}
Long Short-Term Memory networks (LSTMs) is a specialized type of recurrent neural network (RNN) designed to handle sequence data and capture long-term dependencies \cite{Hochreiter_LSTM}. Each LSTM cell uses a gating mechanism—comprising input, forget, and output gates—to regulate the flow of information, making it effective for tasks like time-series prediction. 


 We use the 10-fold cross validation and random search to optimize the models including SVM, Random Forest and XGBoost.
The CNN model consists of two 1D convolutional layers, a fully connected dense layer, and an output layer. The LSTM model is structured with a 100-unit LSTM layer followed by a 50-unit LSTM layer, incorporating Dropout for regularization and ending with a dense layer leading to a single-neuron output.  The CNN and LSTM models are compiled using the Adam optimizer and mean squared error (MSE) loss function.

We selected our base models based on diversity and empirical considerations. These five models come from different domains, have distinct architectures, and utilize varied computational mechanisms, ensuring a broad range of learning capabilities. Given the large number of CFA papers, we also considered commonly used base models and the effectiveness of model fusion. By integrating these five models, we aim to leverage their complementary strengths, enhancing both accuracy and robustness in our predictive framework.

\subsection{Method of Combination}
CFA presents a fresh approach to model fusion and ensemble learning. It capitalizes on computational models that generate score and rank functions, which are derived by the respective scoring systems. The diversity between these systems can be calculated based upon the distance between the proposed Rank-Score characteristic functions using Cognitive Diversity. 

\subsubsection{Score Function, Rank Function, and Rank-Score Function}

Data and information fusion involves combining multiple scoring systems from a variety of sources such as sensor data or decision level systems. CFA can be performed on both the attribute and decision level. It sees each system $A$ as a scoring system, which has a score function $s_A$, and a derived rank function $r_A$, and a rank-score characteristic function $f_A$. Given a set of objects $D=\{d_1, d_2, …, d_n\}$, scores are in the set of real number $\mathbb{R}$, and they are normalized to the range of 0 to 1. A rank function is generated from the scores in the set, with the higher scores corresponding to lower rank numbers. For an ML/AI model $A$, there are score function $s_A$, rank function $r_A$, and RSC function which maps its relationship in the duality of both Euclidean and rank spaces. The RSC function is defined as $f_A(i)=s_A(r_A^{-1}(i))$. \cite{hsu2006combinatorial, hsu2010rank, hsu_kristal_schweikert_2024, Hurley_2021}

The RSC function, defined by Hsu, Shapiro, and Taksa, was used in information retrieval \cite{2002}. The RSC functions of different systems have been used to measure the dissimilarity/diversity between these systems. 

\subsubsection{Cognitive Diversity}

Cognitive diversity (CD) is a measurement to quantify the dissimilarity of a pair of models by computing the area between the RSC functions of the said models \cite{hsu_kristal_hao_schweikert}. It is dataset independent. High CD means that the scoring systems have relatively different ranking and scoring behaviors. The underlying idea is that lower diversity would yield less optimal results because the output would essentially be not much different from the base models. High cognitive diversity is beneficial for combination methods because it allows the systems to complement each other, correcting individual errors and improving overall prediction accuracy.

Cognitive diversity is computed using the following formula:
\begin{equation}
    CD(A, B) = d(f_A, f_B)=\sqrt{\frac{\sum^n_{i=1}(f_A(i)-f_B(i))^2}{n}}
\end{equation}

For a set of $t$ scoring systems, $A_1$, $A_2$, …, $A_t$, the diversity strength $ds(A_j)$ of system $A_j$ is defined as the arithmetic average of cognitive diversity between $A_j$ and other scoring systems \cite{jiang}. It is written as $ds(A_j) = \frac{\sum_{k\neq j}d(A_j,A_k)}{t-1}$ 

The performance strength $p(A_j)$ is determined by the performance of the scoring system $A_j$. The performance strength is assessed based on the designated metrics such as AUROC, accuracy, or precision, depending on the task and the dataset at hand.

\subsubsection{Combination}

When combining the $t$ scoring systems $A_1$, $A_2$, ..., $A_t$, three types of combination strategies are considered \cite{hsu2006combinatorial, hsu_kristal_schweikert_2024, jiang}: 
\begin{enumerate}
    \item average combination (AC)
    \item weighted combination by diversity strength (WCDS)
    \item weighted combination by performance (WCP).
\end{enumerate}

For average score combination (SC) or rank combination (RC), the score function for the score combination $s_{SC}$ and of the rank combination $s_{RC}$ are as follows \cite{hsu_kristal_schweikert_2024, jiang}:

$s_{SC}(d_i) = \frac{\sum^t_{j=1}s_{A_j}(d_i)}{t}, s_{RC}(d_i) = \frac{\sum^t_{j=1}r_{A_j}(d_i)}{t}$

In the case of weighted combination, whether by diversity strength or performance, the score function for score combination and rank combination are calculated as \cite{hsu_kristal_schweikert_2024, jiang}:

$s_{SC}(d_i) = \frac{\sum^t_{j=1}w_j\times s_{A_j}(d_i)}{\sum^t_{j=1}w_j}, s_{RC}(d_i) = \frac{\sum^t_{j=1}\frac{1}{w_j}r_{A_j}(d_i)}{\sum^t_{j=1}\frac{1}{w_j}}$

such that $w_j\in\{p_j, ds_j\}$

\subsection{Methodology Workflow}
\begin{figure}[ht]
    \centering
    \includegraphics[width=0.9\linewidth]{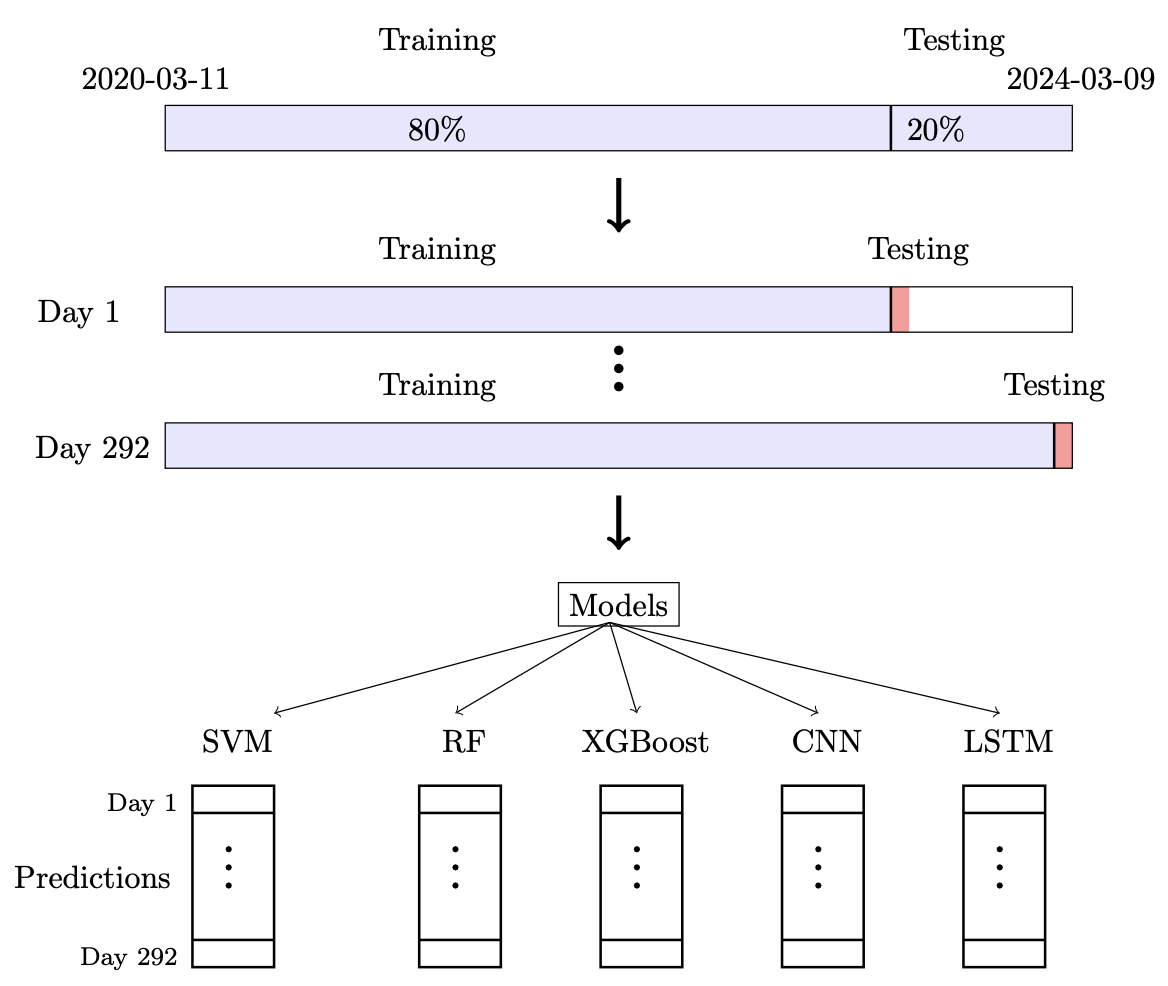}    \vspace{-1em}
    \caption{Phase I of methodology workflow}
    \label{phase1}
\end{figure}

Figure \ref{phase1} illustrates the first phase of our methodology. The dataset starts from March 11, 2020, to March 9, 2024, encapsulating the entirety of the COVID-19 pandemic. To prepare for analysis, we divided the dataset into training and test sets using an 80:20 split. For time series data, employing a sliding window approach is crucial due to the sequential nature of the data, where past values influence future trends. This method ensures that the model learns temporal dependencies effectively, preserving the chronological order of information. In this study, the sliding window begins on the dataset's first day and remains fixed for all subsequent predictions, enabling the model to incorporate the entire historical trajectory of the pandemic. Consequently, predicting a single day's Bitcoin price leverages all preceding days' data. As we progress toward predicting more recent dates, the training set expands to include additional days. Following this preparation, the five models described in Section II(B) are applied to forecast daily Bitcoin prices, as shown in Figure \ref{phase1}.

Similar to prior research efforts in Bitcoin price prediction, we train machine learning models on a designated training set and evaluate their performance on a separate test set. However, our approach distinguishes itself by generating a price distribution for each day's prediction, rather than predicting a single deterministic value. From a statistical perspective, traditional machine learning models focus on estimating the most probable price for a given day, which corresponds to the price with the highest likelihood of occurrence. Our method aims to uncover the range of potential prices that could plausibly occur. By identifying these distributions for each model, we leverage CFA to integrate the predictions from five distinct models, enabling a more robust and comprehensive approach to price forecasting.

\begin{figure}[ht]
    \centering
    \includegraphics[width=0.9\linewidth]{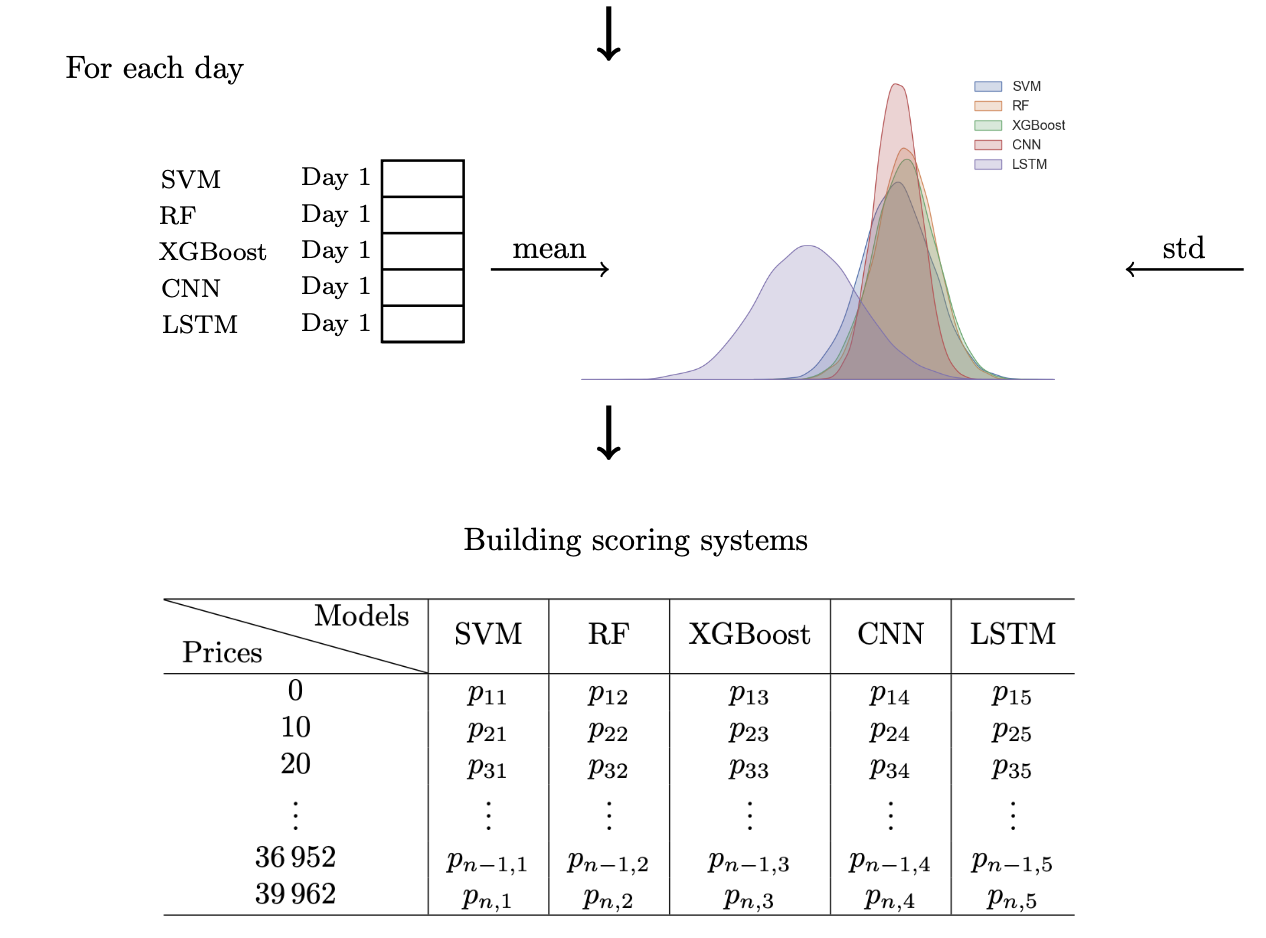}
    \caption{Phase II of methodology workflow}
    \label{phase2}
\end{figure}

As illustrated in Phase II of Figure \ref{phase2}, each daily price prediction from the five base models is expanded into a normal distribution. The predicted price serves as the mean of the distribution, while the standard deviation is derived from the original test set. The standard deviation reflects the variability in a model's predictions, indicating the ability of how different a given model can predict after training on a training set. We suppose, while the predicted prices may vary depending on the training set, the standard deviation remains relatively stable for a given model. Consequently, we use the same standard deviation from the original test set for different days when building a normal distribution.

After generating five normal distributions for each day's prediction, we truncate these distributions at two standard deviations from the mean, capturing approximately $95\%$ of the possible price range. The smallest and largest prices of the five price ranges define the final range. If the lower bound of the final interval is negative, we set the minimum price to zero, as shown in Phase II's table, since negative prices are not meaningful in this context. It is important to note that the precise lower bound is not critical because the final predicted price is determined by the value with the highest probability within the interval after combination. The probabilities derived from the five normal distributions are treated as the scores for each model, while the prices within the intervals serve as the data items. This process establishes five distinct scoring systems, which are then utilized in the subsequent model combination phase.

\begin{figure}[ht]
    \centering
    \includegraphics[width=0.95\linewidth]{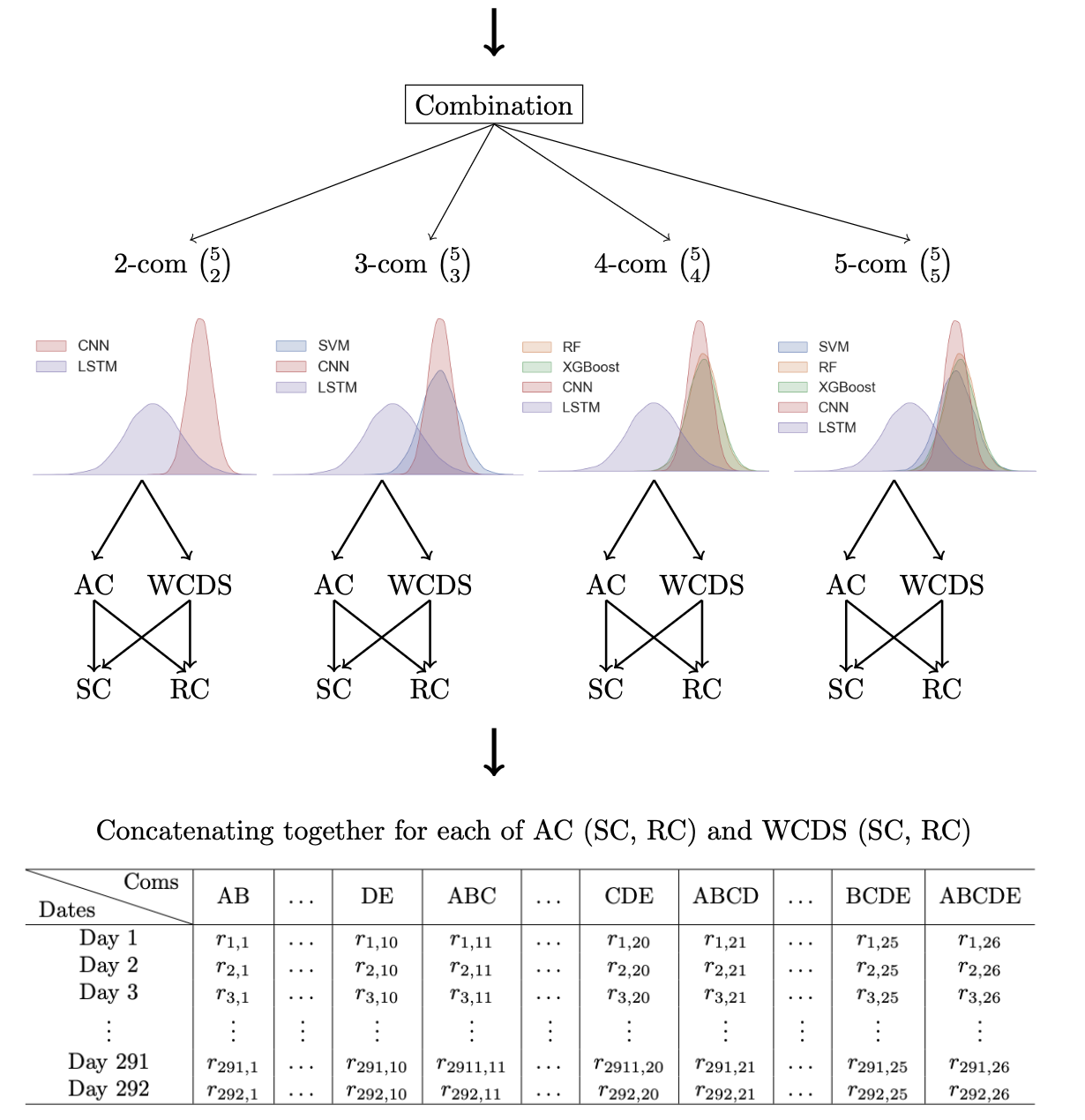}
    \vspace{-1em}
    \caption{Phase III of methodology workflow}
    \label{phase3}
\end{figure}

Phase III in Figure \ref{phase3} describes the process of using CFA, where models are systematically combined to enhance prediction accuracy. The use of combination techniques avoids the existence of extreme points. We first construct combinatorial groups comprising subsets of the five base models, ranging from pairs to all five models together. This process results in a total of 26 unique model groups, calculated as \(\binom{5}{2} + \binom{5}{3} + \binom{5}{4} + \binom{5}{5} = 26\). For each group, we explore four distinct strategies for combining models based on input type (scores or ranks) and weighting approach (average weighting or weighted combination using diversity strength). These strategies include both average score and rank combinations as well as weighted score and rank combinations by diversity strength. As it is not possible to calculate the performance between multiple prices (from a normal distribution) and a single true price, performance weighting is not applied in this paper. Consequently, the 26 model groups, when combined with these four strategies, yield \(26 \times 4 = 104\) unique combination models.

For each combination, we rank the resulting probabilities in descending order to identify the price with the highest probability as the predicted value for a given day. This process generates the table in Phase III, where the columns represent the 26 model groups and the rows correspond to the prediction dates. Since there are four combination strategies, we obtain four analogous tables, each representing a specific approach. These tables form the foundation for selecting the most accurate Bitcoin price prediction for each day by analyzing the corresponding row entries.

\begin{figure}[ht]
    \centering
    \includegraphics[width=1.0\linewidth]{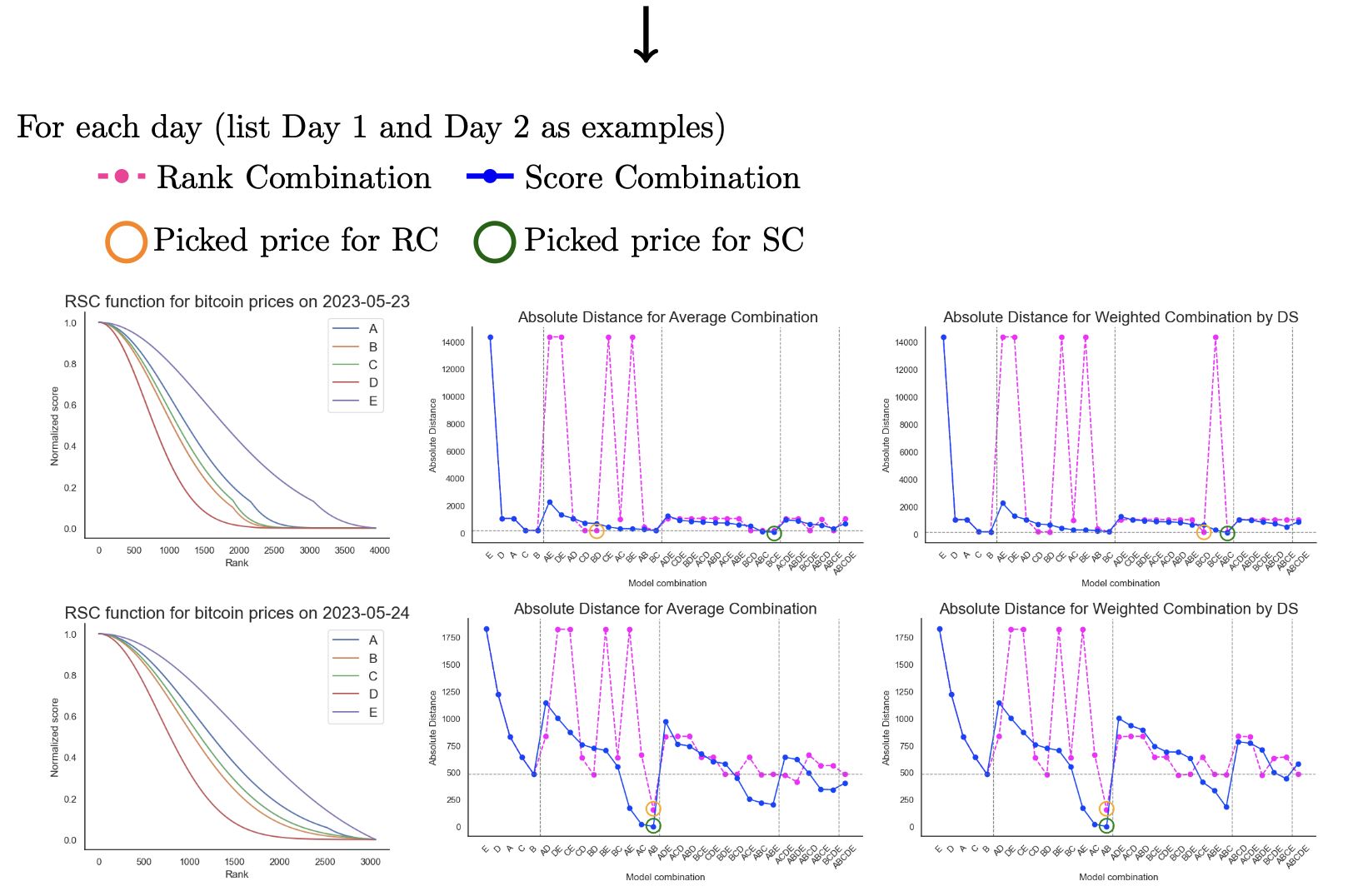}
    \vspace{-1em}
    \caption{Phase IV of methodology workflow}
    \label{phase4}
\end{figure}

In Phase IV of Figure \ref{phase4}, we first determine the absolute distance between the actual price and the predicted price for each model combination. A smaller absolute distance indicates a more accurate predicted price. As illustrated in Figure \ref{phase4}, two scatterplots are provided—one for the average combination method and another for the weighted combination by diversity strength method—each corresponding to daily predictions. Figure \ref{phase4} also includes a rank-score function to represent the diversity among the five base models. Previous research has shown that model combinations outperform individual models when those models exhibit both relatively high predictive performance and large diversity. Consistent with this principle, the five models demonstrate considerable diversity during the first two days of the analysis. Each scatterplot’s x-axis displays five individual models followed by 26 model groups. The score combination method is depicted as a blue solid line, and the rank combination method is depicted as a red dashed line. Our aim is to identify the smallest absolute distance for each combination method, which serves as the final predicted price. In the figures, an orange circle and a green circle highlight the final predicted price for the rank and score combination methods, respectively. If the chosen final predicted price corresponds to a model group rather than an individual model, this outcome indicates that the combination approach has conferred a performance improvement on that particular day. By counting the number of days in which a combination strategy shows improvement, one can assess the effectiveness of that strategy. At the end of Phase IV, we thus determine the final predicted price for each day using four distinct model combination methods.

\section{Results}

\begin{table*}[ht]
    \centering
    \small
    \begin{threeparttable}
    \caption{Performance for individual models and model combinations.}
    \begin{tabular}{p{1.1cm}|c|c|c|c|c||c|c|c|c}
    \hline
   & \multicolumn{5}{c||}{Base Models} 
   & \multicolumn{2}{c}{AC} & \multicolumn{2}{|c}{WCDS} \\
   \cline{2-10}
     & A &B &C &D& E&  SC & RC & SC & RC\\
     \hline
    {\# days$^*$} &-&-&-&-&- &215 &222& 208 & $\boldsymbol{258}$\\
       RMSE  & 1057.34 & 738.21& 966.64 & 2331.47 & 1967.18 & $\boldsymbol{175.22}$ & 289.45& 182.28 & 294.22 \\
        MAPE & $1.86\%$ & $1.20\%$ & $1.55\%$ &$4.91\%$ & $4.68\%$ & $\boldsymbol{0.19\%}$ & $0.40\%$ & $0.22\%$& $0.44\%$ \\
        \hline
    \end{tabular}
    \begin{tablenotes}
    \footnotesize
    \item[*] indicates the number of days that show improvement out of 292 days. 
    \end{tablenotes}
    \label{results}
    \end{threeparttable}
\end{table*}

To forecast Bitcoin prices, we produced a price distribution for each day by using the predicted price from our models as the mean and the standard deviation derived from the test set. CFA was employed to integrate these distributions and determine the optimal predicted price for each combination method. To evaluate the predictive performance of both individual models and combination models, we utilized two widely accepted metrics: Root Mean Squared Error (RMSE) and Mean Absolute Percentage Error (MAPE). The formulas for these metrics are as follows.
\begin{equation*}
    \begin{split}
        RMSE &= \sqrt{\frac{1}{N}\sum_{i=1}^{N} (y_i - \hat y_i)^2},\\
        MAPE &= \frac{1}{N}\sum_{i = 1}^{N} \bigg|\frac{y_i - \hat y_i}{y_i}\bigg| \times 100\%
    \end{split}
\end{equation*}

RMSE and MAPE offer distinct advantages and serve complementary purposes. RMSE, expressed in the same unit as the data (e.g., U.S. dollars), is sensitive to the scale of the dataset. For instance, if Bitcoin prices from the early 2010s—when prices were relatively low—are used, RMSE will naturally be smaller. Consequently, comparing RMSE across datasets with differing scales is not meaningful. In contrast, MAPE is unitless, expressed as a percentage, and remains unaffected by the scale of the dataset. This characteristic allows MAPE to facilitate comparisons across diverse datasets. In this study, RMSE is employed to compare models within our analysis, while MAPE is used to benchmark our models against state-of-the-art models presented in other research studies. The dual use of RMSE and MAPE ensures both intra-study and inter-study comparisons are robust and meaningful.

\begin{table*}[ht]
    \centering
    \small
    \caption{Comparison of the proposed method and previous methods in Bitcoin price prediction.}
    \scalebox{1.1}{
    \begin{tabular}{p{2cm}|p{0.7cm}|p{4cm}|p{4cm}|p{2.5cm}}
    \hline
       Previous Work  & Year & Method & Dataset & Metric \\
    \hline
         Ji, Kim, and Im \cite{ji2019comparative}  & 2019 & Deep neural network (DNN) & Daily Bitcoin price data and blockchain information from 29 
        November 2011 to 31 December 2018 & MAPE: $3.61\%$\\
    \hline
        Ye, Wu, Chen, et al. \cite{ye2022stacking}  & 2021 & Stacking ensemble deep model of 2 base models: LSTM \& GRU & Tweets, transaction data, technical data from September 2017 to January 2021 daily data & MAPE: $4.49\%$\\
    \hline
         Hamayel and Owda \cite{hamayel2021novel} & 2021 & gated recurrent unit (GRU) & Daily Bitcoin price using open, high, low, close features from 22 January 2018 to 30 June 2021 & MAPE: $0.245\%$ \\
    \hline
         Zhang, Li, and Yan \cite{zhang2022empirical} & 2022 & stacking denoising auto-encoders using bootstrap aggregation (SDAE-B) & 9 features of daily bitcoin price data from 29 November 2014 to 30 March 2020 & MAPE: $1.6\%$\\
    \hline
        Kim, Shin, Choi, et al. \cite{kim2022deep} & 2022 & self-attention-based multiple long short-term memory (SAM-LSTM) & Daily Bitcoin price on-chain data from 27 March 2018 to 16 November 2021 & MAPE: $1.33\%$\\
    \hline
        Chen \cite{chen2023analysis}  & 2023 & Random forest & Daily Bitcoin price having 47 features, split into 8 categories from  1 October 2018 to 1 April 2022 & MAPE: $3.29\%$\\
    \hline
        Jin and Li \cite{jin2023cryptocurrency} & 2023 & hybrid prediction model, VMD-AGRU-RESVMD-LSTM, integrating variational mode
        decomposition (VMD), the gated recurrent unit (GRU) and long short-term memory (LSTM) & Daily Bitcoin price from 31 July 2017 to 30 September 2020 & MAPE: $0.394\%$\\
    \hline
        \textbf{Proposed method} & 2025 & Combinatorial fusion analysis (CFA) of 5 base ML/AI models & Daily Bitcoin price from 11 March 2020 to 09 March 2024 & MAPE: $0.19\%$(SCAC) \\
    \hline
    \end{tabular}
    }
    \label{compare}
\end{table*}

In additional, we measure the effectiveness of each combination strategy in CFA using the number of days of improvement which are defined before.
A higher number of improved days indicates that a specific combination strategy is beneficial across a greater portion of the test set.

The performance results are summarized in Table \ref{results}. For the metric of the number of days of improvement, all four combination methods demonstrated improvements for over 200 days out of the 292-day test set, emphasizing the efficacy of the CFA approach. Rank-based strategy outperformed the score-based strategy for both types of combination in terms of the number of improved days. Specifically, the weighted rank combination by diversity strength showed improvements on 258 days, whereas the weighted score combination showed improvements on only 208 days, which marks a significant improvement. These results clearly indicate the robustness of weighted rank combination by diversity strength method.

For the RMSE metric, all model combinations demonstrated superior performance compared to individual models. The lowest RMSE, achieved by average score combination, was 175.22, a substantial improvement over the base models' RMSE of 738.21. Interestingly, while score-based combinations exhibited lower RMSE values than rank-based combinations, rank-based combinations showed improvements on more days. This discrepancy can be attributed to the characteristics of rank-based strategies, which are more likely to yield consistent improvements across a greater number of days but with a slightly smaller magnitude of improvement compared to score-based strategies.

In terms of MAPE, the best performance achieved among the model combinations is $0.19\%$, obtained using the average score combination method. This result is approximately ten times smaller than the MAPE values of any individual base models in Table \ref{results}, highlighting the effectiveness of the combination strategy. Similar to the results observed with RMSE, score combination outperforms rank combination. MAPE is a particularly suitable metric for comparing model performance across datasets, as it provides a standardized measure that is independent of scale. In Table \ref{compare}, we present several reference studies that utilize daily Bitcoin prices as their dataset and evaluate model performance using MAPE. The MAPEs reported in these studies range from $0.245\%$ to $4.49\%$, all of which are higher than the MAPE achieved by our proposed method. Of particular interest in Table \ref{compare} is an advanced ensemble model named VMD-AGRU-RESVMD-LSTM, which integrates variational mode decomposition (VMD), the gated recurrent unit (GRU), long short-term memory (LSTM) neural networks, and attention mechanisms to provide accurate Bitcoin price predictions. Despite its sophisticated architecture, our CFA technique delivers almost twice the performance improvement over their reported result, demonstrating the efficacy and potential of our approach for Bitcoin price prediction.

Average score combination is the best performing methodology in both RMSE and MAPE metric. The underperformance of rank combinations, compared with score combinations, may be attributed to the fact that we have fluctuating diversity strength values for each model. For example, if diversity strength values of the five base models for most of the data items are low, then the rank combinations won't have superior performance than the score combinations \cite{hsu2010rank, 2002}.

\section{Conclusion}

In this article, we proposed a comprehensive approach to predict Bitcoin prices by leveraging diverse features and a moderate set of machine learning models. 
Five distinct machine learning algorithms were employed for raw Bitcoin price predictions. Given the time-series nature of the price dataset, we included Long Short-Term Memory (LSTM) networks to capture temporal dependencies effectively. Moreover, a Convolutional Neural Network (CNN) was utilized to introduce more diversity among base models by leveraging its ability to detect localized patterns within the data. A key innovation of our approach lies in extending Bitcoin price prediction beyond traditional regression frameworks by generating prediction distributions for each day’s price. Through Combinatorial Fusion Analysis, these distributions and their corresponding scoring systems were combined to produce an optimized predicted price. This technique transcends the performance ceiling of traditional models, demonstrating that combining prediction distributions via CFA can push the boundaries of accuracy and reliability in Bitcoin price forecasting.

Performance was evaluated using Root Mean Square Error (RMSE) and Mean Absolute Percentage Error (MAPE), with benchmark results from previous studies included for comparison. Our findings show that CFA-based model combinations consistently outperform individual models, illustrating the power of model fusion in improving predictive accuracy. Furthermore, score-based combinations deliver more precise predictions than rank-based combinations, although rank-based methods still demonstrate higher likelihood for improvement than score-based methods. Notably, our best-performing strategy, average score combination, achieved a MAPE of $0.19\%$, lower than results reported in prior studies. This highlights the superiority and robustness of our approach, as it effectively combines diverse modeling perspectives into a unified and high-performing framework.  

Using sentiment analysis to do price prediction has been a constant area of focus, as people's opinions reflect policy changes and thus can sway the movement of prices \cite{Ouf2024}. Our future work should take into the account of consumer sentiment to more accurately predict the prices. Additionally, we aim to investigate multi-layer CFA frameworks \cite{Hurley_2021} to further refine Bitcoin price predictions and unlock even greater forecasting precision.

A potential limitation of our work is using the test set's standard deviation for normal distribution generation, which may cause data leakage. In future work, a separate validation set can be used to calculate the standard deviation instead.

\printbibliography

\end{document}